# Verificación Formal de Contratos Inteligentes: Una Revisión Sistemática de la Literatura
# Smart Contracts Formal Verification: A Systematic Literature Review

René Davila[1], Everardo Barcenas[1], Rocío Aldeco-Pérez[1].

[1]Instituto de Investigaciones en Matemáticas Aplicadas y en Sistemas - Facultad de Ingeniería, Universidad Nacional Autónoma de México. Circuito Escolar S/N, Ciudad Universitaria, Alcaldía Coyoacán, C.P. 04510, Ciudad de México, México.

photographic_ren@comunidad.unam.mx, ebarcenas@unam.mx, raldeco@unam.mx.

## Abstract

Formal verification entails testing software to ensure it operates as specified. Smart contracts are self-executing contracts with the terms of the agreement directly written into lines of code. They run on blockchain platforms and automatically enforce and execute the terms of an agreement when meeting predefined conditions. However, Smart Contracts, as software models, often contain notable errors in their operation or specifications. This observation prompts us to conduct a focused study examining related works published across various sources. These publications detail specifications, verification tools, and relevant experiments. Subsequently, this survey proposes an alternative formal verification based on description logic.

## Resumen

La verificación formal implica la evaluación del software para garantizar que opere conforme a lo especificado. Los contratos inteligentes son acuerdos autoejecutables cuyos términos están codificados directamente en líneas de programación. Estos contratos operan en plataformas blockchain y ejecutan automáticamente los términos acordados cuando se cumplen condiciones predefinidas. Sin embargo, los contratos inteligentes, en tanto que modelos de software, suelen contener errores significativos en su funcionamiento o especificaciones. Esta observación motiva la realización de un estudio enfocado en analizar trabajos relacionados publicados en diversas fuentes. Estas publicaciones presentan especificaciones, herramientas de verificación y experimentos relevantes. Posteriormente, este estudio propone un enfoque alternativo de verificación formal basado en la lógica descriptiva.

## 1 Introduction

A Smart Contract (SC) is a self-executing contract in which the terms of the agreement are directly written into lines of code. The code and agreements embedded within it exist on a decentralized blockchain network. SCs automatically enforce and execute contractual terms when predefined conditions are met. Since they run on a blockchain, they are decentralized, transparent, and tamper-proof. Smart contracts represent a significant innovation in contract execution, offering increased efficiency, security, and trust in digital agreements [16].

Like any software, SCs can encounter errors during both development and execution. Designing an SC requires careful consideration of all possible scenarios and edge cases. Incomplete or inconsistent logic can result in unexpected behaviors or vulnerabilities. Errors in SC implementation can stem from coding mistakes, including syntax and logic errors, potentially leading to unintended behaviors upon execution [2].

Formal verification is a process that employs mathematical methods to prove or disprove the correctness of a system's design concerning specific formal specifications or properties. Integrating formal verification into SC development enables developers to create more secure, reliable, and efficient contracts, fostering greater confidence and adoption within the blockchain ecosystem [32].

Specifically, this study aims to answer the following Research Questions (RQs): *RQ1: What formal verification methods are applied to SCs in the scientific literature? RQ2: What properties are verified in SCs? RQ3: What experiments and tools are used for verification?*

This review, conducted through a Systematic Literature Review (SLR) [21], explores various formal verification techniques and tools for SCs. Most existing literature focuses on verifying the code implementation and execution of SCs. The SLR highlights the need for verifying SC designs. Therefore, after presenting the background and relevant findings of the SLR, this study proposes a formal verification approach for SC designs using Description Logic (DL) [4], illustrating its application with examples of SCs expressed in DL terms.

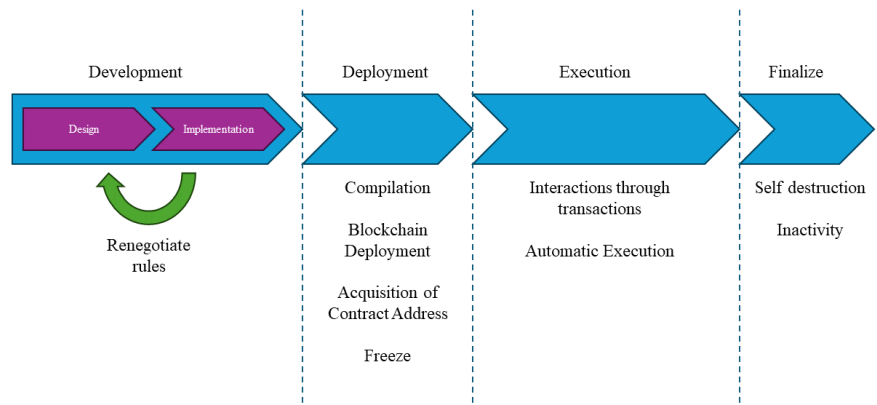

Figure 1. SC Life cycle (Created by the author).

## 2 Background

Blockchain is a decentralized digital ledger technology that enables the secure and transparent recording of transactions across a distributed network of computers. It operates as a continuously growing list of records, known as blocks, which are linked and secured using cryptographic techniques [23].

Each block contains a cryptographic hash of the previous block, a timestamp, and transaction data. Once added to the blockchain, a block becomes virtually immutable, as modifying it retroactively would require altering all subsequent blocks and obtaining consensus from most of the network [23].

Smart contracts are self-executing agreements in which the terms are directly encoded into lines of code. These contracts are stored and executed on blockchain networks, such as Ethereum [8], and automatically enforce the agreed-upon terms when predefined conditions are met [16].

The terms and conditions of a smart contract are written in code by the parties involved or developers. These terms may include rules, obligations, and actions that the contract will execute when specific conditions are met. Once deployed on the blockchain, the smart contract becomes immutable and is stored across multiple nodes within the network. This ensures transparency, security, and resistance to tampering. When the predefined conditions encoded in the contract are met (e.g., a specific date is reached, or a payment is received), the contract automatically executes the specified actions [16].

### 2.1 Smart Contract Life Cycle

Table 1. Selected Literature on the development and execution phases of the SC life cycle.

| Smart Contract Verification | | | |
|---|---|---|---|
| SC Life Cycle | Verification Method | Tool | Reference |
| Development – Design | Model Checking | PROMELA & SPIN | [5] |
| Development – Implementation | Model Checking | NuSMV | [26] |
| Development – Implementation | Model Checking | BIP & SMC | [1] |
| Development – Implementation | Model Checking | VeriSolid | [24] |
| Execution | Model Checking | MCMAS | [25] |
| Execution | Theorem Proving | Isabelle/HOL | [3] |
| Execution | Runtime Verification | LARVA | [12] |
| Execution | Symbolic Execution | MAIAN | [28] |
| Execution | Abstract Interpretation | MadMax | [14] |
| Execution | Fuzzing | ContractFuzzer | [19] |
| Execution | Theorem Proving | Coq/ConCert | [31] |
| Execution | Vulnerability Detection | GPTSCAN | [34] |

Figure 1 illustrates the life cycle of a smart contract (SC) [30], which consists of several key phases, from creation to execution and eventual termination.

*Design:* This phase defines the contract's logic, conditions, functions, and necessary variables. It also involves a detailed analysis of requirements and use cases [30].

*Implementation:* The contract is implemented using a specific programming language for SCs, such as Solidity (for Ethereum [8]) or other languages depending on the blockchain platform. This phase highlights the adaptability and flexibility of SC development, making it an engaging aspect for blockchain enthusiasts [30].

If stakeholders determine that modifications to the SC's rules are necessary before deployment, the rules are renegotiated during this phase [30].

*Interactions through transactions:* Users interact with the contract by sending transactions to its address. These interactions can trigger contract functions and modify its state [18].

*Automatic Execution:* When predefined conditions are met, the contract automatically executes certain functions. Its internal logic ensures compliance with predefined rules without requiring human intervention [18].

**2.2 Formal Methods**

Formal verification is a method used in computer science and engineering to rigorously prove or verify the correctness of hardware and software systems. It employs mathematical techniques to demonstrate that a system satisfies specified requirements or properties [11]. Table 2 presents selected literature on descriptive logic.

Table 2. Description Logics Selected Literature.

| Description Logic Verification | | | |
|---|---|---|---|
| Property | Tool | SC Verification | Reference |
| Consistency | SpecCC | No | [33] |
| Consistency | KL-ONE | No | [17] |
| Consistency | Semantic Web | No | [10] |

Below are key formal verification techniques commonly used in related works included in this Systematic Literature Review (SLR):

*Model Checking.* A formal verification method that systematically explores all possible states of a system model to determine whether specific properties hold true or whether undesired behaviors may occur [6].

*Theorem Proving.* A formal verification approach that uses logical deduction to prove the correctness of mathematical statements or theorems. It derives theorems from axioms and existing theorems through a sequence of logical steps [15].

*Runtime Verification.* A method that monitors and analyzes a system's behavior during execution to ensure adherence to specified properties or requirements. Unlike other formal verification techniques, which typically involve static analysis before deployment, runtime verification operates dynamically during execution [13].

*Symbolic Execution.* A program analysis technique used for software testing, debugging, and verification. It explores program behavior by executing it symbolically rather than using concrete input values, representing variables and expressions as symbolic formulas instead [20].

*Abstract Interpretation.* A formal framework for the static analysis of programs. It approximates and reasons about program behavior without necessarily analyzing all possible executions or concrete values. Instead, it operates over abstract representations of program states, enabling efficient analysis while providing insights into program properties [9].

*Fuzzing.* Also known as fuzz testing or fuzz analysis, this technique is used to discover vulnerabilities or software bugs by providing invalid, unexpected, or random inputs. The goal of fuzzing is to identify potential security flaws, crashes, or unintended behaviors that could be exploited by attackers or lead to software failures [22].

## 3 Review Methodology

To incorporate diverse perspectives and strengthen the research in this study, we followed the systematic literature review (SLR) methodology outlined in [21], while also considers additional recommendations from the same source. The structured mapping process we employed consisted of the following steps: first, we formulated a search query ("Formal Verification" AND "Smart Contracts") to capture the intersection of these research domains. Next, we addressed the research questions (RQs) to guide our investigation. We then identified relevant information sources for the literature search. Finally, we filtered relevant documents from these sources based on predefined inclusion and exclusion criteria.

Our SLR [21] aims to develop a comprehensive understanding of program synthesis components, explore the mechanics of automatic program generation, and identify areas requiring further investigation to fully grasp the context and optimal aspects of synthesized program creation.

*RQ1: What formal verification methods are applied to SCs in the scientific literature?* Several information sources were selected to search for relevant documents related to formal verification and SCs.

To ensure a focused study, we applied specific inclusion and exclusion criteria to refine the data screening process. The inclusion criteria consisted of journal articles and conference proceedings, while the exclusion criteria ruled out articles from books, courses, magazines, standards, and Ph.D. theses. Based on these guidelines, the search process was further refined by filtering publications from 2018 to 2024. Additionally, we analyzed the titles, abstracts, and keywords of journal articles and conference papers that met the inclusion criteria to extract relevant information.

A summary of the method's application is presented in the results. A total of 44 documents were retrieved from the designated data sources, out of which 12 papers were directly relevant to the research questions. The findings are subsequently organized according to the research questions that guided the systematic literature mapping.

## 3.1 Formal Verification

This subsection addresses *RQ2: What properties are verified in SCs?* and *RQ3: What experiments and tools are used for verification?* using the selected literature. Table 1 presents the related documents, categorized according to the formal verification methods applied. The literature analyzed in this section corresponds to the development and execution phases of the SC life cycle (Figure 1).

## 3.2 Formal Verification at Development Phase

Continuing with the response to RQ2 and RQ3, this subsection presents the relevant literature categorized by the formal verification methods applied, specifically highlighting results from the implementation stage. The selected literature for this analysis corresponds to the development phase of the SC life cycle (Figure 1, Section 2.2).

*Formal Modeling and Verification of Smart Contracts.* The approach described in Formal Modeling and Verification of Smart Contracts [5] integrates formal methods into smart contract development to minimize errors and reduce costs. This methodology generates a standardized smart contract template by applying formal modeling techniques, which represent contracts using tuple structures and finite state machine concepts.

```
pragma solidity ^0.8.17;

contract EtherWallet {

    address payable public owner;

    constructor() {

        owner = payable(msg.sender);}

    receive() external payable() {}

    function withdraw (uint _amount) external {

        require (msg.sender == owner, "caller is not owner");

        payable (msg.sender).transfer(_amount);}

        function getBalance() external view returns(uint) {

        return address(this).balance;}}
```

Figure 2. EtherWallet Example.

To validate this approach, the authors employed PROMELA (Process Meta Language) to model a shopping smart contract (SSC) and used the SPIN model checker to verify the correctness of the model and assess essential properties. This verification process ensures that the contract adheres to predefined specifications, reducing the likelihood of logic errors before deployment.

*Model-Checking of Smart Contracts.* The model consists of three main layers: a kernel layer capturing blockchain behavior (specifically Ethereum), an application layer modeling smart contracts (including translation rules from Solidity to NuSMV), and an environment layer defining the application execution framework. To assess contract behavior, expected properties must be expressed in temporal logic, covering aspects such as safety, fairness, reachability, and real-time attributes. If a property fails, model-checking provides a counterexample, allowing for the identification of defects. This method is adaptable and can be applied to various Ethereum applications. In a case study, the model-checking approach successfully verified four out of five properties, with the unsatisfied property accompanied by a counterexample from the model checker [26].

## 4 Description logic verification

DLs offer a way to represent relationships among entities within a specific domain. DLs ontologies are constructed using three fundamental elements: concepts, which denote groups of entities; roles, which denote binary relationships between entities; and individual names, which identify specific entities within the domain. Those familiar with first-order logic will recognize these as unary predicates (for concepts), binary predicates (for roles), and constants (for individual names) [29].

A KB consists of two main components: the TBox and the ABox. The TBox defines the terminology, which is the vocabulary used within a specific application domain, while the ABox contains assertions about named individuals using this vocabulary [4]. Therefore, the vocabulary includes concepts representing sets of individuals and roles denoting binary relationships between individuals. DL systems allow users to define complex descriptions of concepts and roles in addition to atomic concepts and roles (concept and role names) [29].

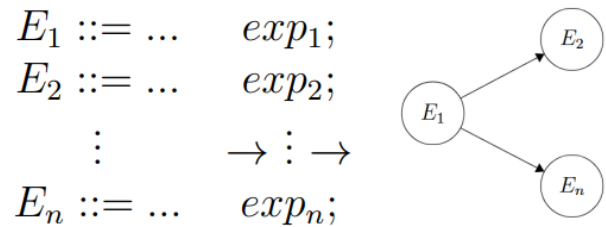

$$\begin{aligned} E_1 &::= \ldots \quad exp_1; \\ E_2 &::= \ldots \quad exp_2; \\ &\vdots \quad\quad\quad \to \vdots \to \\ E_n &::= \ldots \quad exp_n; \end{aligned}$$

Figure 3. Schematic diagram of the algorithm.

After showing the theoretical bases of DLs and before reviewing an example of SC expressed in DLs, a selection of DL literature related to our verification proposal is presented in Table 2.

*Selected literature on formal verification with DLs*

### 4.1 Smart Contracts case study and Verification Proposal

Using the theoretical concepts of DL presented previously, the following examples in the Solidity language for Ethereum illustrate the expressiveness of DL for SCs.

*Example – EtherWallet*

The given Solidity code (see Figure 2) defines a simple Ethereum [8] wallet contract that allows the owner to deposit, withdraw funds, and check the balance. The critical components of the contract include the owner, the constructor to set the owner, a function to withdraw funds, and a function to get the balance.

In Description Logic, we typically deal with concepts (unary predicates) and roles (binary predicates). Here is how we can map the Solidity code to DL:

Concepts: *EtherWallet* - represents the contract itself; *Owner* - represents the owner of the wallet.

Roles: *owns* - a role indicating that the owner has control over the wallet; *canWithdraw* - a role indicating that the owner can perform withdrawals; *hasBalance* - a role indicating the balance of the wallet.

Individuals: *owner* - an individual representing the owner's address; *wallet* - an individual representing the wallet instance.

$EtherWallet(wallet)$ asserts that *wallet* is an instance of the *EtherWallet* concept.

$Owner(owner) \land owns(owner, wallet)$ asserts that owner is an instance of the *Owner* concept and *owner* owns the *wallet*.

$\forall x (canWithdraw(x, wallet) \rightarrow Owner(x) \land owns(x, wallet))$ states that if any entity *x* can withdraw from the wallet, then *x* must be the owner and must own the wallet.

$\exists x\, hasBalance(x, wallet)$ asserts that the wallet has some balance associated with it.

The previous sections have motivated us to develop a Smart Contract verification system, where defining the concept of consistency is considered a crucial first step. To achieve this, a grammar that incorporates multiple programming languages is first constructed. This grammar serves as the foundation for an algorithm designed to build finite state machines, which are intended to represent the execution flow of a Smart Contract. The goal is to identify consistency issues within the rules implemented in programming languages such as Solidity or Go (see Figure 3).

Once these inconsistencies are identified, Description Logics will be employed to verify the logical correctness of the Smart Contract rules. As previously discussed, this approach allows for the expression of various elements present in the Smart Contract code.

## 5 Conclusions

After analyzing a wide range of literature on the Formal Verification of Smart Contracts, Model checking is the most prevalent method for verifying smart contract properties, especially during the development phase. Various formal verification tools are necessary to assess different aspects of smart contracts at the development phase. They ensure that defined properties and rules in a smart contract do not result in contradictions or logical errors. The study suggests exploring formal verification frameworks based on description logic to address inconsistencies in smart contracts during the design stage. It is an area of opportunity to explore.

This work was supported by the Mexican Secretary SECIHTI (1006953), UNAM-PAPIIT TA101723 and UNAM-PAPIIT IA104724.

## References

[1] T. Abdellatif and K. Brousmiche. Formal verification of smart contracts based on users and blockchain behaviors models. In IFIP NTMS International Workshop on Blockchains and Smart Contracts (BSC) (2018).

[2] M. Almakhour, L. Sliman, A. Samhat and A. Mellouk. Verification of smart contracts: A survey. Pervasive and Mobile Computing 67 (2020).

[3] S. Amani, M. Bégel, M. Bortin, and M. Staples. Towards verifying Ethereum smart contract in isabelle/hol. In Proceedings of the 7[th] ACM SIGPLAN International Conference on Certified Programs and Proofs (2018), CPP, Association for Computing Machinery.

[4] F. Baader, D. Calvanese, D. McGuinness, D. Nardi, and P. Patel-Schneider. The Description Logic Handbook: Theory, Implementation and Applications, 2ed. Cambridge University Press, 2007.

[5] X. Bai, Z. Cheng, Z. Duan and K. Hu. Formal modeling and verification of smart contracts. In Proceedings of the 2018 7th International Conference on Software and Computer Applications (2018), ICSCA '18, Association for Computing Machinery.

[6] C. Baier and J. Katoen. Principles of Model Checking (Representation and Mind Series). The MIT Press, 2008.

[7] M. Browne, E. Clarke and O. Grümberg. Characterizing finite kripke structures in propositional temporal logic. Theoretical Computer Science 59, 1 (1988).

[8] V. Buterin. A next-generation smart contract and decentralized application platform – white paper. Ethereum Project (2014).

[9] P. Cousot. Formal verification by abstract interpretation. In NASA Formal Methods (2012), A. E. Goodloe and S. Person, Eds., Springer Berlin Heidelberg.

[10] L. Dragone and R. Rosati. Checking e-service consistency using description logics. In IEEE International Conference on Services Computing (2007).

[11] R. Drechsler. Formal system verification. Springer, 2018.

[12] J. Ellul and G. Pace. Runtime verification of Ethereum smart contracts. In 2018 14th European Dependable Computing Conference (EDCC) (2018).

[13] Y. Falcone, K. Havelund and G. Reger. A Tutorial on Runtime Verification. In Engineering Dependable Software Systems, M. Broy, D. Peled, and G. Klaus, Eds., vol. 34 of NATO Science for Peace and Security Series – D: Information and Communication Security. IOS Press, 2013.

[14] N. Grech, M. Kong, A. Jurisevic, L. Brent, B. Scholz and Y. Smaragdakis. Madmax: surviving out-of-gas conditions in Ethereum smart contracts. Proc. ACM Program. Lang. 2, OOPSLA (2018).

[15] J. Harrison. Theorem proving for verification. In Computer Aided Verification (2008), A. Gupta and S. Malik, Eds., Springer.

[16] M. Herlihy. Blockchains from a distributed computing perspective. Commun. ACM 62, 2 (2019).

[17] P. Hors and M. Rousset. Consistency of structured knowledge: A formal framework based on description logics. Expert Systems with Applications 8, 3 (1995). European Verification and Validation of Knowledge-Based Systems.

R. Davila et al. / Abstraction & Application 50 (2025) 46 – 56	55

[18] Y. Huang, Y. Bian, R. Li, J. Zhao and P. Shi. Smart contract security: A software lifecycle perspective. IEEE Access 7 (2019).

[19] B. Jiang, Y. Liu and W. Chan. Contractfuzzer: fuzzing smart contracts for vulnerability detection. In Proceedings of the 33rd ACM/IEEE International Conference on Automated Software Engineering (2018), ASE '18, Association for Computing Machinery.

[20] J. King. Symbolic execution and program testing. Commun. ACM 19, 7 (1976).

[21] B. Kitchenham, O. Pearl, D. Budgen, M. Turner, J. Bailey and S. Linkman. Systematic literature reviews in software engineering – a systematic literature review. Information and Software Technology 51, 1 (2009).

[22] G. Klees, A. Ruef, B. Cooper, S. Wei and M. Hicks. Evaluating fuzz testing. In Proceedings of the 2018 ACM SIGSAC conference on computer and communications security.

[23] J. Liu and Z. Liu. A survey on security verification of blockchain smart contracts. IEEE Access 7 (2019).

[24] A. Mavridou, A. Laszka, E. Stachtiari and A. Dubey. Verisolid: Correct-by-design smart contracts for Ethereum. In Financial Cryptography and Data Security: 23$^{rd}$ International Conference, FC 2019, Revised Selected Papers, Springer-Verlag.

[25] W. Nam and H. Kil. Formal verification of blockchain smart contracts via atl model checking. IEEE Access 10 (2022).

[26] Z. Nehaï, P. Piriou and F. Daumas. Model-checking of smart contracts. In 2018 IEEE International Conference on Internet of Things (iThings) and IEEE Green Computing and Communications (GreenCom) and IEEE Cyber, Physical and Social Computing (CP-SCom) and IEEE Smart Data (SmartData).

[27] D. Neider and I. Gavran. Learning linear temporal properties. In 2018 Formal Methods in Computer Aided Design (FMCAD) IEEE.

[28] I. Nikolic, A. Kolluri, I. Sergey, P. Saxena and A. Hobor. Finding the greedy, prodigal, and suicidal contracts at scale. In Proceedings of the 34$^{th}$ Annual Computer Security Applications Conference (2018), ACSAC '18, Association for Computing Machinery.

[29] S. Rudolph. Foundations of description logics. In Reasoning Web International Summer School. Springer, 2011.

[30] C. Sillaber and B. Waltl. Life cycle of smart contracts in blockchain ecosystems. Datenschutz und Datensicherheit-DuD 41, 8 (2017).

[31] D. Sorensen. Towards Formally Specifying and Verifying Smart Contract Upgrades in Coq. In 5th International Workshop on Formal Methods for Blockchains (FMBC 2024). Open Access Series in Informatics (OASIcs), Volume 118, Schloss Dagstuhl – Leibniz-Zentrum für Informatik.

[32] Y. Wang, S. Lahiri, S. Chen, R. Pan, I. Dillig, C. Born, I. Naseer and K. Ferles. Formal verification of workflow policies of smart contracts in azure blockchain. In Verified Software. Theories, Tools, and Experiments (2020), S. Chakraborty and J. A. Navas, Eds., Springer International Publishing.

[33] R. Yan, C. Cheng and Y. Chai. Formal consistency checking over specifications in natural languages. In 2015 Design, Automation & Test in Europe Conference & Exhibition (DATE).

[34] S. Yuqiang, W. Daoyuan, X. Yue, L. Han, W. Haijun, X. Zhengzi, X. Xiaofei and L. Yang. 2024. GPTScan: Detecting Logic Vulnerabilities in Smart Contracts by Combining GPT with Program Analysis. In Proceedings of the IEEE/ACM 46th International Conference on Software Engineering (ICSE '24). Association for Computing Machinery, Article 166.